\documentstyle[multicol,prl,aps,epsfig]{revtex}

\begin{document}

\twocolumn[\hsize\textwidth\columnwidth\hsize\csname @twocolumnfalse\endcsname

\title{Boundary effects on one-particle spectra of Luttinger liquids }

\draft
\author {K.\ Sch\"onhammer$^1$, V.\ Meden$^2$, 
 W.\ Metzner$^2$, U.\ Schollw\"ock$^3$, and O.\ Gunnarsson$^4$}
\address{$^1$Institut f\"ur Theoretische Physik, Universit\"at
  G\"ottingen, Bunsenstr.\ 9, D-37073 G\"ottingen, Germany\\
$^2$Institut f\"ur Theoretische Physik C, RWTH Aachen, D-52056 Aachen,
Germany\\
$^3$Sektion Physik, Universit\"at M\"unchen, Theresienstr.\ 37,
D-80333 M\"unchen, Germany \\ 
$^4$Max-Planck-Institut f\"ur Festk\"orperforschung, D-70506
Stuttgart, Germany}

\date{December 22, 1999}
\maketitle

\begin{abstract}
We calculate one-particle spectra for a variety of 
models of Luttinger liquids with open boundary 
conditions. For the repulsive Hubbard model the 
spectral weight close to the boundary is enhanced in 
a large energy range around the chemical potential. 
A power law suppression, previously predicted by 
bosonization, only occurs after a crossover at energies
very close to the chemical potential. Our comparison 
with exact spectra shows that the 
effects of boundaries can partly be understood within the 
Hartree-Fock approximation.

\end{abstract}
\vskip 2pc]
\vskip 0.1 truein
\narrowtext

Interacting fermions in one spatial dimension 
do not obey Fermi liquid theory.
It took about thirty 
years to
fully understand the generic low-energy physics of one-dimensional
(1D) fermions with repulsive interaction, now called {\em Luttinger 
liquid} (LL) behavior\cite{Voit}. For various correlation functions
LL theory predicts anomalous power laws
with interaction dependent exponents\cite{Voit}.  
Several systems were studied 
using different experimental techniques in
order to verify these predictions. Photoemission or transport
measurements seem to be most promising\cite{Voit,Zwick,Grioni}.
 While the basic understanding of LL behavior emerged in
the study of 1D systems with 
{\em periodic boundary conditions} (PBC), it
is obvious that in order to describe realistic geometries more
realistic boundary conditions have to be used. Therefore hard walls,
usually called {\em ``open'' boundary conditions}
(OBC) were
studied recently\cite{FG,EMJ,WVP}. 
For the translational invariant system standard
renormalization group (RG) arguments can be used\cite{So}
to show that for repulsive 
interactions
the $2k_F$-scattering part 
(usually called ``$g_1$-interaction'')  
scales to zero and the Tomonaga-Luttinger model (TLM) describes the
generic low energy physics. In some of the previous publications 
on the open boundary
problem\cite{FG,EMJ} it was {\it tacitly assumed} that similar RG arguments
can also be applied to the system without translational
invariance. Therefore a quadratic form in boson operators was used to
describe the electron-electron interaction. 
Then
it is straightforward to calculate correlation
functions\cite{FG,EMJ}. 
It was shown that the
local spectral density $\rho(x,\omega)$ entering the description 
of angular integrated
photoemission is strongly modified near the endpoints of a 1D
chain. The algebraic {\it vanishing} of the spectral density with frequency
$\omega$ close to the chemical potential $\mu$ was found to be governed by a
coupling dependent {\em boundary exponent} $\alpha_B$ which, 
for repulsive interaction,
is larger than the bulk exponent\cite{FG,EMJ}. This result was 
used recently in the
interpretation of photoemission
experiments\cite{Zwick,Grioni}. 
In Ref.\ \cite{WVP} $\alpha_B$ has been studied for a variety 
of integrable models using results of the Bethe ansatz (BA) and assuming 
conformal invariance. 

Thus far only specialized methods (bosonization, the BA 
and conformal field theory) have been used to study the one-particle 
properties of LL's with boundaries and assumptions were
necessary to justify the applicability of these methods.
Furthermore these methods do 
only provide the exponent which 
characterizes $\rho(x,\omega)$ at {\it asymptotically small} energies.
In this letter we study several models for LL's 
with OBC using bosonization and the density matrix 
renormalization group (DMRG) method. 
By comparison with results obtained from these (numerically) 
{\it exact} methods we will show that the low 
energy behavior of $\rho(x,\omega)$ can partly be understood
within the (non-self-consistent) {\it Hartree-Fock} (HF)
{\it approximation}  for the self-energy, which already gives power law
behavior. This is a very surprising observation 
as the HF approximation for the case of PBC does not capture any of
the LL features. 
For the models considered we
will explicitly demonstrate that $\alpha_B$ can be expressed in 
terms of the bulk
LL parameter $K_{\rho}$, $\alpha_B =(K^{-1}_{\rho}-1)/z$, where $z=1$ in
the spinless case and $z=2$ for spin $1/2$-fermions\cite{FG,EMJ,WVP}.
For the TLM and the lattice model of spinless fermions (SF) with nearest 
neighbor interaction the exponent 
$\alpha_B^{\rm HF}$ found in HF agrees to leading order in the 
coupling with $\alpha_B$ and for small couplings
there is {\it quantitative} agreement between 
the exact local Green's function and the HF approximation.
We will show that for the Hubbard model (HM) with 
small or moderate 
Coulomb repulsion $U$ the spectral weight 
at the endpoint of the chain is  {\it enhanced} in a large energy 
range close to $\mu$. A {\it crossover} to a power law decrease occurs in 
a surprisingly small energy range around $\mu$.  
This crossover behavior cannot be
obtained from bosonization or conformal field theory 
due to the limitation to asymptotically small energies. 
Because of the ``flowing
$g_1$-coupling''\cite{So} there is 
only {\it qualitative} agreement between DMRG and the 
HF approximation in case of the HM.
Besides their relevance for LL's with OBC our results are important
for the investigation of LL's with PBC including 
impurities, since such models are expected to 
scale to chains with OBC\cite{KF}.

We first present a
careful discussion of the interaction term of the Hamiltonian in the
basis of the one-particle eigenstates $\varphi_n (x) = \sqrt{2/L} 
\sin{ (k_n x)}$ with $k_n = n\pi/L$,  $n \in {\rm I \! N}$ 
for OBC, a continuum model, and an interaction with a spatial range $R$.
Here $L$ denotes the length of the chain.
Note that the $k_n$ do {\it not} have the meaning of momenta. 
In terms
of the creation and annihilation operators $a^{(\dagger)}_n$ of
the eigenstates $\varphi_n$ the two-body interaction
takes the usual form\cite{spinfootnote}
\begin{eqnarray*}
\hat V = \frac{1}{2}\sum_{n,n',m,m'}v_{nmn'm'} \, a^\dagger_n a^\dagger_m
a^{}_{m'} a^{}_{n'} , 
\end{eqnarray*}
with the matrix elements $v_{nmn'm'}$. If we express
products of the sine-functions $\varphi_n$
 in terms of cosine-functions they read
\begin{eqnarray}
&& v_{nmn'm'} = 
\left[ F\left(k_n-k_{n'},k_m-k_{m'}\right) \right. \nonumber \\
&& - F\left(k_n-k_{n'},k_m + k_{m'}\right)  
- F\left(k_n+k_{n'},k_m - k_{m'}\right) \nonumber \\ 
&& \left. + F\left(k_n + k_{n'},k_m + k_{m'}\right) \right]/L,
\label{eqn2}
\end{eqnarray}
with
\begin{eqnarray*}
F(q,q')  = 
\frac{1}{L} 
\int^L_0 dx \int^L_0 dx' \cos{( qx)} V(x-x')\cos{(q'x')}.
\end{eqnarray*}
Due to the fact that $k_n = n\pi/L$ (and {\em not} $2\pi n/L$) in
addition to terms involving Kronecker deltas a 
finite size correction appears
for {\it non-local} interactions
\begin{equation}
F(q,q') =  \tilde{V}(q)/2 \, \left( \delta_{q,q'} +
  \delta_{q,-q'}\right) + g(q,q')/L,
\label{eqn3}
\end{equation}
where $\tilde V(q)$ is the Fourier transform of the interaction
$V(x)$. 
The correction term $g(q,q')/L$ does not contribute to 
$\alpha_B$\cite{FG} and we thus neglect this term. 
For an interaction which is long range in real space, considered by 
Tomonaga for PBC,
$\tilde V(k)$ is different from zero for 
$|k|< k_c$, where the cut-off $k_c=1/R$ is much smaller than 
the Fermi momentum $k_F = n_F \pi/L$. Then only the first term on 
the right hand side (rhs) of
Eq.\ (\ref{eqn2}) is important for the low energy physics and $\hat V$ can be
written as a bilinear form in boson operators.
Using bosonization correlation functions can then 
be calculated as discussed in Ref.\ \cite{FG} yielding $\alpha_B =
(K^{-1}_{\rho}-1)/z$ with $K_{\rho} = [1 + z\tilde
V(0)/(\pi v_F)]^{-1/2}$ and the Fermi velocity $v_F$. 
The full $\omega$ dependence of
$\rho(x,\omega)$ for the TLM has been presented in Fig.\ 1 of 
Ref.\ \cite{EMJ}.

\begin{figure}[thb]
\begin{center}
\vspace{-0.0cm}
\leavevmode
\epsfxsize6.5cm
\epsffile{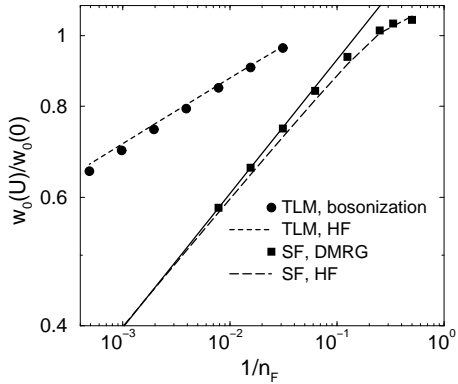}
\caption{Spectral weight at $\mu$
  and close to the boundary. The circles show the exact results 
  for the spinless TLM at $x=3R$ with a long range 
  interaction $\tilde{V}(q) = U \Theta(k_c-|q|)$,
  $U/(2 \pi v_F) =0.1$ and $k_F = 4 k_c$. The dashed 
  line presents the HF
  approximation. The squares show the exact results 
  for the lattice model of SF at site $i=1$ 
  for $n_F/N=0.25$ and $U=1$. The
  respective HF approximation is given by the long dashed line. 
  The solid line displays a power law with exponent 
  $\alpha_B=0.1838$. The error of the DMRG data 
  is of the order of the symbol size. }
\label{fig1}
\end{center}
\vspace{-0.0cm}
\end{figure}

In contrast to the bulk exponent,
which, for weak coupling, is quadratic in the 
interaction, $\alpha_B$ has a contribution {\it linear} in the 
interaction. Thus signs of the non-analytic behavior
of $\rho(x,\omega)$ can already be obtained using 
the HF self-energy. The 
delta terms on the rhs of Eq.\ (\ref{eqn3}) yield 
\begin{eqnarray}
&& \left( \Sigma^{\rm HF} \right)_{kk'}  =  
\delta_{kk'} \left\{ \delta \mu-
 \frac{1}{2 L} \sum_{k_{1}\le k_{F}} \left[ \tilde V(k-k_1) +
    \tilde V(k+k_1) \right]  \right\} \nonumber\\ 
 && + 
 \frac{1}{2 L} \left\{ z\tilde V(k+k')- \tilde V
  \left( [k-k']/2 \right) \right\}
f\left( [k+k']/2\right) \label{eqn5}  \\
&& - 
 \frac{1}{2 L} \left\{ z \tilde V(k-k') - \tilde V
  \left([k+k']/2 \right) \right\} f\left(|k-k'|/2\right) \nonumber,
\end{eqnarray}
where $f(k_m)$ is the
Fermi function, which is equal to one
if $m$ is integer and $m \le n_F$ and zero 
otherwise, and $\delta \mu =z\tilde V(0) n_F/L$.
The calculation of $\rho^{\rm HF}$ requires a matrix
inversion. 
The leading non-analytic behavior in the HF approximation can be traced 
to the term in  $\Sigma^{\rm HF}$ which is proportional to
$f([k+k']/2)$. 
To {\it leading order}
in $\tilde{V}$ the term  in the second line of Eq.\ (\ref{eqn5}) gives a
contribution to $\rho^{\rm HF}$  proportional to 
$[ \tilde{V} (0)         - z \tilde{V}(2 k_F)] 
 \, \log |\omega - \mu ^{\rm HF}|$. 
For a long range repulsive interaction $\tilde{V}(2 k_F) = 0$ 
and the prefactor of the logarithm 
is positive. This indicates the suppression of spectral weight close
to $\mu ^{\rm HF}$. Note that the logarithmic divergence 
is {\em not} due to a singular 
frequency behavior of $\Sigma^{\rm HF}$, as it occurs in second order
perturbation theory for PBC, but emerges in the process
of the matrix inversion. For finite systems it is numerically 
straightforward to perform the matrix inversion and calculate 
$\rho^{\rm HF}$.
Instead of comparing $\rho^{\rm HF}$ and the exact spectral function 
$\rho$ obtained from bosonization as a function of $\omega$ 
we have studied the spectral weight at the chemical potential and 
position $x$, denoted    
$w_0(x,n_F;\tilde{V})$, for a given $k_F$ as a function of $n_F$.
The ratio $w_0(x,n_F;\tilde{V})/ w_0(x,n_F;0)$ as a function of 
$1/n_F$ displays the same kind of power law behavior as does 
$\rho(x,\omega)$ as a function of $\omega$. 
For the lattice models considered later we will only be able to exactly
determine the spectral weight at the
chemical potential on a lattice site $i$ as a function of 
the system size.
In Fig.\ \ref{fig1} we 
compare the $1/n_F$ dependence of $w_0(x,n_F;\tilde{V})/ w_0(x,n_F;0)$ 
for the exact solution and the HF approximation. 
Already the HF approximation displays power law behavior, which shows
that the leading logarithmic divergence of $\rho^{\rm HF}$ discussed
above can be resummed to give a power law. A detailed
study shows that $\alpha_B^{\rm HF}=\tilde{V}(0)/(2 \pi v_{\rm HF})$ where 
$v_{\rm HF} = v_F + \tilde{V}(0)/ (2 \pi)$ is the Fermi velocity in
the HF approximation and thus $\alpha_B^{\rm HF}$ and 
$\alpha_B$ do agree up to leading order in   
$\tilde{V}(0)/( 2 \pi v_F) $.
Quantitative agreement between
exact results and HF can be reached for 
$ \tilde{V}(0) / (2 \pi v_F) \ll 1$.

For the lattice models (SF, HM) 
considered next even the case with OBC 
can be solved exactly by the BA, but 
as for PBC not much about correlation functions 
can be learned directly from the BA. 
Information about boundary exponents can be obtained 
if conformal invariance is assumed\cite{WVP}.
First we will discuss the local spectral function at site 
$i=1$ in a lattice model of SF with lattice constant 
$a \equiv 1$, hopping matrix 
element $t \equiv 1$, nearest neighbor interaction $U$, and OBC. We have
performed a DMRG study\cite{dmrg} for chains of up to $N=512$ sites 
calculating matrix elements
$w_0(n_F;U)=|\langle E^{n_F-1}_0|c_1|E^{n_F}_0 \rangle|^2$, i.\ e.\ the 
spectral weight at the chemical potential
and the boundary site. Here $|E^{n_F}_0 \rangle$
denotes the exact $n_F$-particle groundstate  and $c_1$  destroys a
fermion at site $1$. Typical data for $w_0(n_F;U)/w_0(n_F;0)$ 
at quarter filling and for $U=1$ are shown in Fig.\ \ref{fig1}. 
For these parameters 
$K_{\rho}=0.8447$ as can be found in Ref.\ \cite{qin}. 
For large $n_F$ the numerical data nicely follow the solid line, which is 
proportional to a power law with exponent $\alpha_B=0.1838$ 
and we can thus conclude, that for
SF the suppression is described  by $\alpha_B$. 

The HF self-energy for SF is best studied in the site
representation. Due to the
OBC the site occupancies $\langle n_i\rangle$ and the renormalization
of the hopping amplitudes $\langle c_{i+1}^{\dag} c_i  \rangle$
depend on $i$ and in order to obtain the site diagonal Green's 
function one numerically has to invert a tridiagonal 
matrix. Fig.\ \ref{fig1} presents HF data for $w_0(n_F;U)/w_0(n_F;0)$
and the same para\-meters as above. The HF approximation
shows power law behavior with exponent $\alpha_B^{\rm HF}=\tilde{V}_{\rm eff}/
(2 \pi v_{\rm HF})$ which has the same form as for the TLM if one replaces 
$\tilde{V}(0)$ by the effective interaction 
$\tilde{V}_{\rm eff} = \tilde{V}(0)-\tilde{V}(2k_F) = 
2 U [1-\cos{(2k_F)}]$.
Again
$\alpha_B$ and $\alpha_B^{\rm HF}$ do agree up to order 
$\tilde{V}_{\rm eff}/(2 \pi v_F) $ and  
both curves show quantitative agreement for 
$\tilde{V}_{\rm eff}/(2 \pi v_F)  \ll 1$.

\begin{figure}[tb]
\begin{center}
\vspace{-0.0cm}
\leavevmode
\epsfxsize6.5cm
\epsffile{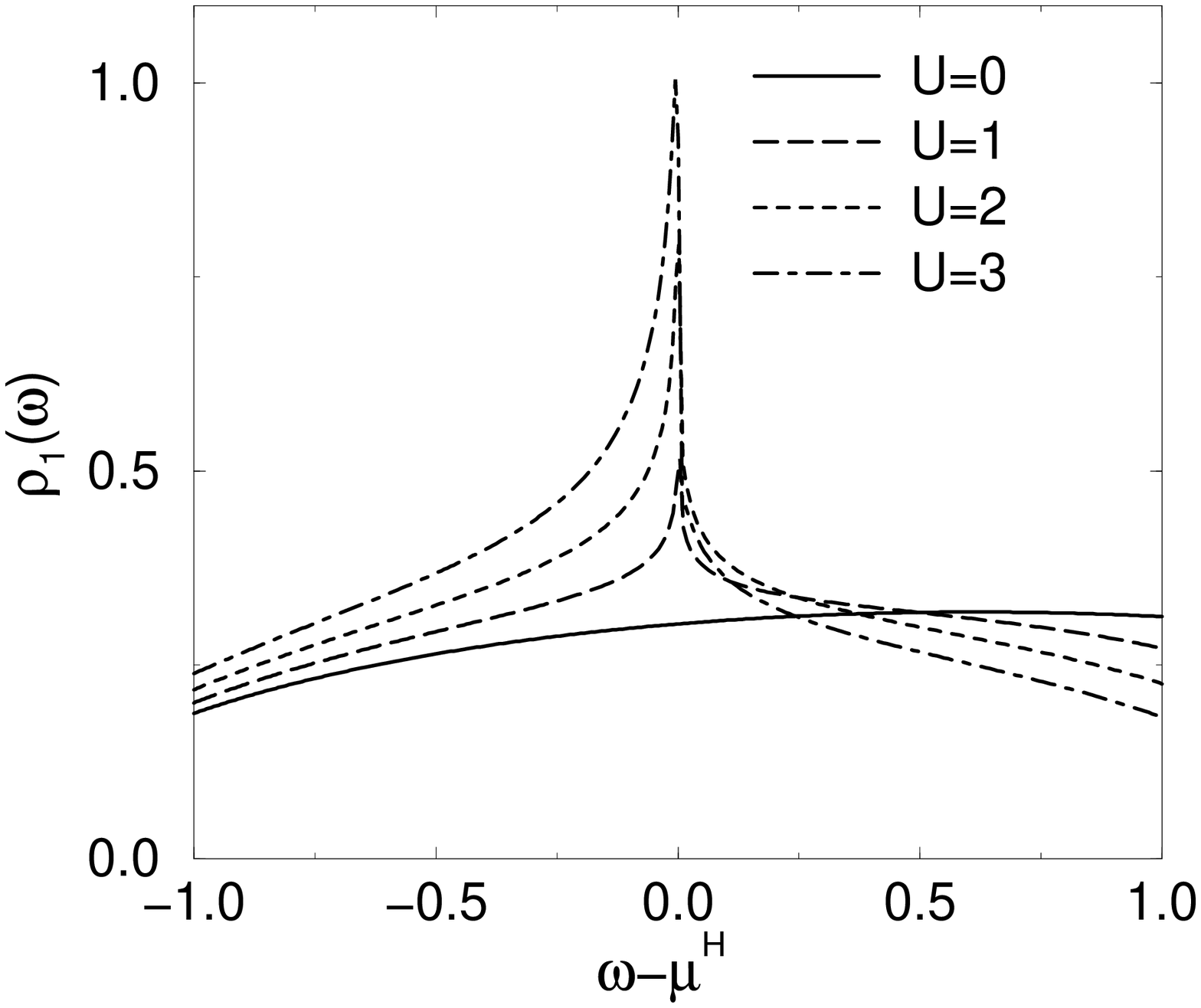}
\caption{Spectral density at the boundary site for the HM 
with $N=800$ sites and $n_F=320$ in the Hartree approximation
and for different values of $U$.} 
\label{fig2}
\vspace{-0.0cm}
\leavevmode
\epsfxsize6.5cm
\epsffile{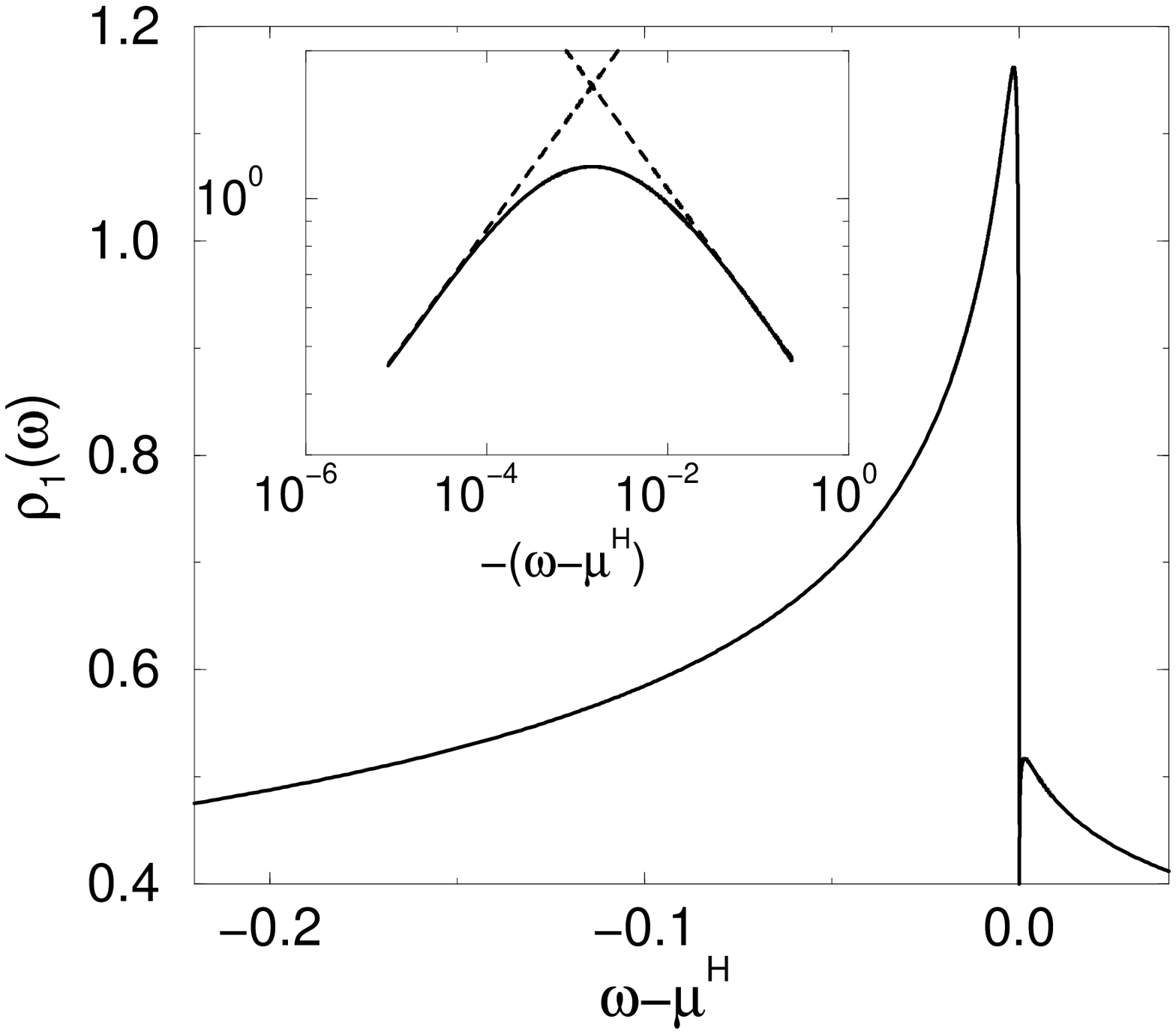}
\caption{The same as in Fig.\ 2, but 
for $U=3$, $N=10^6$, and $n_F=4 \cdot 10^5$. 
The inset shows the spectral weight for the occupied
states $\omega < \mu^{H} $ on a log-log scale
(solid line). The dashed lines display power laws with exponents $\pm
U/(2 \pi v_F)$.}
\label{fig3}
\end{center}
\vspace{-0.0cm}
\end{figure}

Next we will consider the HM and first discuss the results 
of the Hartree approximation (the Fock term vanishes).
To obtain the site diagonal Green's 
function one again has to invert a tridiagonal 
matrix which we have done  numerically for systems of up to 
$N=10^6$ lattice sites.
The corresponding spectral function $\rho_i(\omega)$ at the
boundary site $i=1$ is shown in Fig.\ \ref{fig2} for various values of 
the onsite interaction $U$ and fixed filling $n_F/N=0.4$. 
We have connected the individual weights of the finite system to a
continuous line.
In contrast to the expectation from the general statement in 
previous work\cite{EMJ} the spectral weight near $\mu^{H}$ 
is strongly {\it enhanced.}
This could have been expected already from Eq.\ (\ref{eqn5}).
The prefactor  $\tilde{V} (0)  -z \tilde{V}(2 k_F)$
of the leading logarithmic correction to $\rho^{\rm HF}$ is given by
$(1-z)\tilde V(0)$ for a $k$-independent delta interaction, i.e.
in a model with spin it
has the {\em opposite } sign as in the case of a long range interaction.
As long as $[U/ (2 \pi v_F)] \log{(N)}  \ll 1$
this indicates a logarithmic increase of the weight close to $\mu^H$.
If we go to larger values of
 $ [U/ (2 \pi v_F)] \log{(N)}$, i.e.\ larger $N$ (or $U$), 
as in Fig.\ \ref{fig3}, we observe a {\em
  crossover} to a power law {\em decay} for energies extremely close
to $\mu^{H} $. 
It can be shown 
that the crossover occurs at energies which are 
exponentially (in $-1/U$) close to $\mu^{H}$.
In Fig.\ \ref{fig2} the crossover cannot be observed as the
energy resolution $\Delta \omega \sim 1/N$ is too low. 
In the inset of Fig.\ \ref{fig3} the spectral weight of the
occupied states is shown on a
log-log scale. The increase of weight close to 
$\mu^{H} $ is given by a power law with exponent $- U/(2 \pi v_F)$ 
which is cut off at the crossover energy. This shows that the
logarithmic divergence obtained in leading order in
$U$ can be resummed to produce a power 


\noindent law in a large energy
range close to $\mu^H$. The exponent of the subsequent power law
suppression is given by $U/(2 \pi v_F)$.

\begin{figure}[b]
\begin{center}
\vspace{-0.0cm}
\leavevmode
\epsfxsize6.5cm
\epsffile{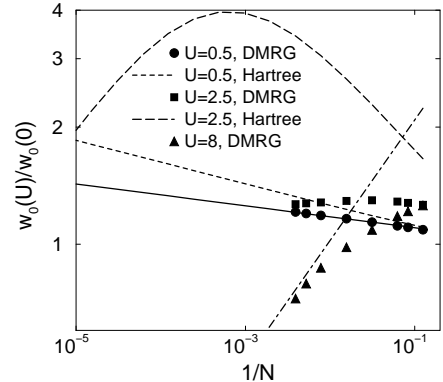}
\caption{Spectral weight at $\mu$
  and the boundary site for the quarter filled HM. The 
  symbols show the results obtained from DMRG. 
  The dashed curves display the weight in the
  Hartree approximation. The solid line shows a power law 
  proportional to $(1/N)^{- \alpha_B}$ for $U=0.5$ and the
  dashed-dotted line is proportional to $(1/N)^{\alpha_B}$ 
  for $U=8$. The error of the DMRG data 
  is of the order of the  symbol size.} 
\label{fig4}
\end{center}
\vspace{-0.0cm}
\end{figure}

In order to demonstrate that the crossover behavior found in the Hartree 
approximation is present also in the exact spectral function 
we again have calculated matrix elements
$w_0(n_F;U)$ using DMRG for chains of up 
to $N=256$ sites. 
In Fig.\ \ref{fig4} $w_0(n_F;U)/ w_0(n_F;0)$ is presented 
for $n_F/N=0.25$, $U=0.5, 2.5$ and $U=8$. 
For small $N$ and $U$ the DMRG 
and the Hartree results do approach each other as expected.
A detailed study 
shows that the difference between the weights for fixed $N$ goes like 
$U^2$.
If one goes from small to large $N$ the Hartree result for 
$U=2.5$ diplays the power law increase with
exponent $-U/(2 \pi v_F)$, the crossover, and the subsequent decrease 
with exponent $U/(2 \pi v_F)$ similar to $\rho_1(\omega)$ as a function of
$\omega$.
For $U=0.5$ only the increase can be
seen as the crossover occurs at much longer
chain length not shown in Fig.\ \ref{fig4}. The exact weight 
obtained 
from the DMRG also
increases for increasing $N$. From the $U=2.5$ data it is clear
that the crossover occurs at smaller 
$N$ and the maximum is less pronounced compared to the Hartree result.  
For $U=8$ the weight already decreases even for the smallest system
sizes available. The dashed-dotted line displays 
a power law with
exponent $\alpha_B$.
The bulk LL parameter $K_{\rho}$ for a given $U$ and $n_F/N$
can be found in Ref.\ \cite{Schulz}. 
The DMRG data are consistent with the conclusion that the
final power law suppression near the boundary is given by
the boundary exponent  $\alpha_B $\cite{FG,EMJ,WVP}.
The increase of weight in the $U=0.5$ DMRG data  
follows a line which is proportional to 
a power law with exponent $-\alpha_B$. 
The leading behavior of the boundary 
exponent is $\alpha_B \approx U/(4 \pi v_F) $\cite{missprint}  
which is  {\it one-half}
of the exponent $U/(2 \pi v_F)$ found in the Hartree 
approximation. This kind of discrepancy between exponents obtained 
in perturbation theory and 
the leading behavior of exact exponents is known from the 
HM with PBC and occurs because a fixed point is reached only
when backscattering has scaled to zero\cite{So}. 
For $U=2.5$ the crossover length is too small to observe a well defined 
power law increase.
We can conlcude that analogous to the Hartree solution 
the DMRG results are consistent with 
an increase of spectral weight 
for small  $[U/ (2 \pi v_F)] \log (N)$, a crossover
and a subsequent power law decrease.

In conclusion for the models studied 
we have explicitly confirmed the expression for the  
boundary exponent $\alpha_B$ in terms of the bulk LL parameter 
$K_{\rho}$ obtained from bosonization and the BA. 
This exponent characterizes the one-particle 
spectral weight  
for asymptotically small energies and close to the boundary.
Investigating the spectral function at higher energies
we have found a quite unexpected behavior for the HM. For 
small and moderate positive $U$ the spectral weight 
close to the boundary first increases
when the energy 
approaches the chemical potential from below and only in a narrow energy range 
the power law decrease with exponent $\alpha_B$ sets in.
Whether LL behavior has convincingly been    
demonstrated in photoemission
experiments is still a matter of debate\cite{Grioni}. 
Nonetheless the above 
scenario has to be taken into consideration in any
discussion of the influence which boundaries (open ends or 
impurities) might have on the photoemission spectra of 
quasi-one-dimensional conductors.
We have furthermore demonstrated that the influence of boundaries 
on one-particle spectra of LL's can partly be understood within 
the HF approximation.

This work  has been supported by the SFB 341
of the Deutsche Forschungsgemeinschaft (V.\ M.\ and W.\ M.). 
Part of the calculations 
were carried out on a T3E of the Forschungszentrum J\"ulich.


\begin{thebibliography}{*}
\bibitem{spinfootnote} For simplification we have suppressed the spin 
index and have assumed a spin independent interaction.
\bibitem{missprint} The leading behavior of $K_{\rho}$ is given by
  $K_{\rho} = 1- U/(2 \pi v_F) $. Note the misprint in Eq.\
  (8) of Ref.\ [12].
\end{thebibliography}
\end{document}